\documentclass[sigconf]{acmart}

\AtBeginDocument{%
  \providecommand\BibTeX{{%
    \normalfont B\kern-0.5em{\scshape i\kern-0.25em b}\kern-0.8em\TeX}}}

\acmConference[KDD `22]{SIGKDD '22: ACM SIGKDD International Conference on Knowledge Discovery and Data Mining}{August 14-18, 2022}{Washington, DC, USA}
\usepackage{todonotes} % for comments
\usepackage{amsmath}
\usepackage{afterpage}

\newcommand\blankpage{%
    \null
    \thispagestyle{empty}%
    \addtocounter{page}{-1}%
    \newpage}
\begin{document}

\title{Flurry: a Fast Framework for Reproducible Multi-layered Provenance Graph Representation Learning}

\author{Maya Kapoor}
\affiliation{\institution{Parsons Corporation}
             \institution{UNC Charlotte}
             \country{}}
             \email{mkapoor1@uncc.edu}
\author{Joshua Melton}
\affiliation{\institution{UNC Charlotte}
            \country{}}
             \email{jmelto30@uncc.edu}
\author{Michael Ridenhour}
\affiliation{\institution{UNC Charlotte}
            \country{}}
             \email{mridenh7@uncc.edu}
\author{Mahalavanya Sriram}
\affiliation{\institution{UNC Charlotte}
            \country{}}
            \email{msriram2@uncc.edu}
\author{Thomas Moyer}
\affiliation{\institution{UNC Charlotte}
            \country{}}
            \email{tmoyer2@uncc.edu}
\author{Siddharth Krishnan}
\affiliation{\institution{UNC Charlotte}
            \country{}}
            \email{skrishnan@uncc.edu}

\begin{abstract}
Complex heterogeneous dynamic networks like knowledge graphs are powerful constructs that can be used in modeling data provenance from computer systems.
%From a security perspective, the graph model enables causality analysis and tracing for analyzing cyber attacks, but there is currently not a pipeline or framework for turning system executions and provenance into nodes and edges and into machine learning data points and feature vectors.
From a security perspective, these attributed graphs enable causality analysis and tracing for analyzing a myriad of cyberattacks. However, there is a paucity in systematic development of pipelines that transform system executions and provenance into usable graph representations for machine learning tasks.
%This lack of instrumentation is an inhibitor to scientific development and experimental reproducibility in the field of provenance graph machine learning.
This lack of instrumentation severely inhibits scientific advancement in provenance graph machine learning by hindering reproducibility and limiting the availability of data that are critical for techniques like graph neural networks.
%To fulfill this need, we present presents Flurry, an end-to-end data pipeline which recreates cyber attacks, captures provenance data from these attacks at multiple system and application layers, converts audit logs from these attacks into data provenance graphs, and conveys these network representations to selected or hand-designed graph deep learning models for anomaly detection and causal analysis in real, resilient systems.
To fulfill this need, we present \textsc{Flurry}, an end-to-end data pipeline which simulates cyberattacks, captures provenance data from these attacks at multiple system and application layers, converts audit logs from these attacks into data provenance graphs, and incorporates this data with a framework for training deep neural models that supports preconfigured or custom-designed models for analysis in real-world resilient systems.
We showcase this pipeline by processing data from multiple system attacks and performing anomaly detection via graph classification using current benchmark graph representational learning frameworks.
\textsc{Flurry} provides a fast, customizable, extensible, and transparent solution for providing this much needed data to cybersecurity professionals.
%We evaluate the results of analyzing provenance graphs generated from the attacks and benign behavior offered in our fast provenance generation framework using some of the models in the pipeline.
%Results show significant capability in detecting anomalous execution graphs, as well as the novel ability to recreate datasets and rerun experiments in a fast, efficient, and transparent manner.
\end{abstract}




\maketitle

\section{Introduction}

The lineage or record of data modification, access, and usage, known as \textit{data provenance}, has been a long-standing tool for resilient cybersecurity systems to use in attack tracing. Data may be sourced from multiple layers of a host, including the kernel or operating system level, the application or data layers, the browser layer, and the physical or virtualized file system. A single host may produce several gigabytes of log data a day even at a coarse level of provenance capture~\cite{camflow}. With multiple host systems generating intranet data over long periods of time, it is evident that data provenance management and analysis is a big data problem.

Resilient systems use graph representations of data provenance in intrusion and anomaly detection in order to perform attack tracing and to find causal dependencies to form mitigation strategies used against all kinds of attacks. Provenance graphs, often represented as heterogeneous networks, are comprised of system entities and objects interconnected by a variety of interactions generated from agents who affect changes in the system. Representing system activity using knowledge graphs effectively captures the complex and multi-faceted nature of system activity and lends itself to structural analysis by graph machine learning algorithms, causal inference, and time series analysis over dynamically generated networks. Unsupervised learning from the structure of the network representation has allowed security specialists to detect even zero-day attacks that have not been previously seen and therefore existing supervised machine learning techniques lack the training data to effectively identify such attack vectors~\cite{sigl, unicorn}. Additionally, the ability of graph machine learning techniques to effectively scale up to web-scale graphs, coupled with the use of graph summarization techniques~\cite{unicorn, streamspot} has allowed for ``low and slow" attack patterns like Advanced Persistent Threats (APTs) to be detected~\cite{prov-gem} despite the sheer amount of data gathered over long periods of time, which has thwarted traditional IDSes.

The advantages of graph machine learning applied to provenance graphs for anomaly detection have inspired a growing, intersectional research community of cybersecurity specialists, systems engineers, data scientists, and machine learning experts. Each member of this community has specialized knowledge of different parts of the non-trivial process of logging data, engineering the data into graphs, and learning from this graph-structured data. Currently, there is not a data pipeline for automatically producing data from cyberattacks, generating provenance graphs, and testing graph machine learning algorithms on the data; rather, this process must be done manually by experts from across the fields. This can be challenging, for example, for data scientists who may not have access to cyberattack data or the specialization to generate and capture this kind of data themselves. On the other hand, cybersecurity specialists may not have the graph machine learning expertise to effectively apply the latest machine learning approaches to the system provenance data they have on hand.

In order to bridge the gap between these communities, enable automatic data generation, and optimize the learning process, we propose \textsc{Flurry}, a fast provenance framework for graph generation and analysis. \textsc{Flurry} is a virtual machine environment equipped with provenance capture and graph generation tools. As opposed to learning on static datasets, \textsc{Flurry} dynamically executes automated cyberattacks and converts these logs into provenance graphs ingestible by state-of-the-art graph machine learning tools. With the \textsc{Flurry} system, researchers in provenance graph representation learning are no longer required to perform their work in separate silos. The process from attack simulation to anomaly detection is transparent, explainable, and readily extensible for use with current techniques and for future development of novel techniques for automated provenance analysis. Researchers may additionally use our provenance generation system with their own machine learning models for plug-and-play functionality to test new methods of representation learning.
\bigbreak
The \textsc{Flurry} system provides the following contributions:
\begin{itemize}
\item{An end-to-end pipeline which can take a system execution, model it as a multi-layer provenance graph, and use graph representation learning to detect anomalies in that execution,}
\item{Dynamic, reproducible data sets of web-based injection attacks and brute force password attacks,}
\item{Visualized execution of cyberattacks at the push of a button for full clarity into our experiments and simplicity for future researchers,}
\item{Concepts for multi-layered data provenance across kernel and application layers, bridging the semantic gap between user and kernel space,}
\item{A plug-and-play framework for graph learning models to analyze complex provenance graphs. Furthermore, our framework provides a readily usable sandbox to test graph machine learning models designed to analyze whole system provenance graphs,}
\item{A platform to bring together researchers from the machine learning community and the systems community.}
\end{itemize}

\section{Preliminaries}

\subsection{Data Provenance}
Provenance refers to the lineage of an object over time. Specifically, provenance is a record of data origination, modification, and interactions. Provenance originated in the context of art history, but has been widely adapted even to computer system data. This kind of data provenance records the history of system data over time, but also importantly captures relationships among data. These associations can provide credibility to the data and be used to establish trust. Furthermore, this trustworthiness makes data reusable for both scientific research and root cause analysis. Provenance records may be replayed or analyzed in a security context to determine potential cyber attacks and mitigate those threats as a response.

Due to its relational properties, provenance data can be well-represented as a provenance graph. The W3C Incubator Group has defined the PROV-DM model in order to represent provenance data as directed, acyclic graphs. For node types, data provenance can be recorded as entities, which are real, physical or digital concepts; activities, which are interactions between or among entities; or agents, which can act upon entities. PROV-DM also defines several relation types between these nodes~\cite{w3c-prov}. Provenance graphs are \textit{directed}, meaning that an entity for example acts upon an activity in a directional relationship. The graphs are also \textit{acyclic} due to the time series of actions. The graphs which are produced by provenance data are by nature \textit{heterogeneous}, as nodes will be different files, processes, inodes, and more diverse node types which will have their own unique features. Provenance produces \textit{multigraphs} which have multiple relationships between nodes of varying edge types, making the graph problem both rich with information and uniquely challenging for analysis.

\begin{figure} [ht]
  \includegraphics[width=\linewidth]{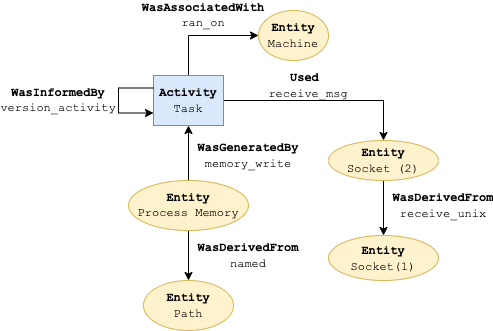}
  \caption{Sample provenance graph from Google Chrome web browsing activity.}
  \label{fig:chromegraph}
\end{figure}

Figure~\ref{fig:chromegraph} presents a high-level abstraction of provenance gathered from the Google Chrome web browser running on the \textsc{Flurry} system. In the diagram, the Google Chrome web browsing service (task) was generated by the process memory which was derived from the binary path. This task is running on the host machine and receiving messages from a socket. Furthermore, the task is informing itself of versioning activity as it is running. This depiction represents only a small portion of the gigabytes of data that are generated as part of the provenance process over lengthy periods of time.

\subsection{Graph Representation Learning}
For data provenance, graph representation learning may be used to detect anomalies in execution graphs. Executions of anomalous behavior (for example, a cross-site scripting attack) will generate provenance subgraphs which differ structurally and in their hidden representations from benign execution graphs (for example, web requests from a browser). Graph machine learning can aid cyber security analysts in detecting and adapting to unforeseen attacks in resilient systems.

We represent this multi-relational provenance data as a heterogenous graph $G = (V, E, O, T, R)$. This directed graph contains a nodeset $V$, an edgeset $E$, a set of node types $O$, a set of edgetypes $T$, and a set of canonical relation types $R$. Each node $v \in V$ maps to a node type $o_v \in O$ and each edge $(u, v) \in E$ maps to an edge type $t_{u, v} \in T$. Each canonical relation type $r \in R$ maps to a tuple $(o_u, t_{u, v}, o_v)$. The aim of the graph representation learning component of this framework is to learn a function which maps a provenance graph $G$ to an expressive, low-dimensional vector representation, or embedding, $z$, where distance between embeddings correlates to similarity between the graphs. In this work graph embeddings are used to train a classifier to detect graphs containing anomalous behavior.

\section{Related Work}

The collection, storage, and summarization of data provenance have been well-studied in the security research community, and representational learning on heterogeneous provenance graphs has recently become a prominent field in the machine learning community~\cite{multinet}. Despite long-standing research efforts in their respective communities, the intersection of data provenance and graph representational learning lacks many foundational frameworks that provide abundant public datasets, encourage reproducibility, and enable the use of graph neural networks for system provenance data. Recent efforts in developing graph-based techniques that employ neural networks for automated analysis of system provenance data make the need for such frameworks increasingly clear.

\subsection{Graph Representation Learning}
Graph representation learning seeks to generate low-dimensional feature vectors (or embeddings) for nodes, edges, or graphs. Traditional methods for generating embeddings for homogeneous graphs, such as DeepWalk~\cite{deepwalk} and node2vec~\cite{node2vec} used random walks on graphs to learn about the local and global neighborhoods in which nodes participate, allowing for generation of node representations that preserve graph structural information. Advances in deep graph neural networks have introduced the message-passing paradigm and convolutional graph embedding approaches (GCNs)~\cite{kipf_gcn, graphsage}. Such GCNs leverage spectral or spatial convolutional filters along with neighborhood aggregation to encode graph structural properties in node embeddings. Graph Attention Networks (GAT)~\cite{gat} further improved on such graph convolutional approaches by introducing node-level attention mechanisms during neighbor aggregation.

Such convolutional approaches were originally limited to homogeneous networks, but in recent years a number of heterogeneous graph embedding frameworks have been proposed to extend graph neural networks to such multi-layered networks. Relational-GCN~\cite{rgcn} extends the idea of GCNs by employing separate convolutional operations for each relation within a knowledge graph and then aggregating node representations across the multiple relations in which each node participates. Heterogeneous Attention Networks (HAN) similarly extend the idea of GAT to heterogeneous networks by employing multiple convolutional filters defined over specified metapaths in a graph. To consolidate the various representations of nodes from multiple metapaths, HAN employs semantic attention to learn an optimal aggregation of the multiple embeddings. Graph isomorphism network (GIN)~\cite{gin} explores the expressive power of GCN models for graph classification. The authors prove that graph convolutional approaches can at most equal the Weisfeiler-Lehman algorithm for graph isomorphism and that sum aggregation is more expressive of the characteristic of multi-sets when compared with mean or max aggregators. In addition to these works, other recent frameworks extend the ideas of random walks on heterogeneous graphs~\cite{metapath2vec,multinet}, and spectral approaches~\cite{fame} for use with graph neural networks that are effective techniques for unsupervised learning on heterogeneous graphs.

\subsection{Deep Learning on Provenance Graphs}
\subsubsection{Streamspot}
Manzoor et al propose StreamSpot~\cite{streamspot} as a clustering-based anomaly detection approach on heterogeneous streaming graphs. They analyze browsing data and downloading software in an effort to identify drive-by download attacks as anomalous behavior.
\subsubsection{SIGL}
SIGL~\cite{sigl} uses a word2vec-based node embedding approach in conjunction with a long short-term memory network to capture long-term dependencies in the provenance graph. Instances that fail to reach a minimum threshold of reconstruction loss in the decoder stage are determined to be anomalous software installations.
\subsubsection{Unicorn}
Similar to StreamSpot, Unicorn~\cite{unicorn} uses a graph sketching technique based on similarity hashing to reduce graph size to a fixed space. For classification, the Unicorn system employs a version of the Weisfeiler-Lehman subtree graph kernel algorithm. They use clustering-based anomaly detection algorithms to aggregate similar graph hashes and identify anomalous provenance graph signatures.
\subsubsection{PROV-GEm}
In PROV-GEm~\cite{prov-gem}, the authors propose a graph embedding framework based on graph convolutional networks coupled with relational self-attention to generate informative representations of provenance graphs. This combination allows PROV-GEm to encode the multi-faceted relationships and graph structural properties captured in heterogeneous data provenance graphs. PROV-GEm conducts anomaly detection experiments using the datasets published by the authors of StreamSpot and Unicorn, outperforming the aforementioned methods on graph anomaly detection and suggesting great potential for graph neural network approaches in automated analysis of data provenance graphs.
\bigbreak
While the datasets in these related works are publicly available to researchers, they are static in nature. System provenance data cannot be easily reproduced with certainty since system specifications and attack scenarios may differ, and potential researchers must rely on the sparse documentation of previously published provenance data generation frameworks. In this work, we propose a paradigm shift to \textit{dynamic} data collection for provenance graph analysis. With the \textsc{Flurry} system, researchers can design their own benign and attack scenarios with support for a wider variety of attack types than in previous research. Raw system data is dynamically created, and the logs are converted to graphs using our extensible software module. This ensures that researchers also have access to the original system logs, providing a better representation of real-world attack scenarios than curated, pre-processed datasets~\cite{streamspot-dataset}. \textsc{Flurry}'s intuitive GUI illustrates attack scenarios in the Chrome browser in real time, providing a transparent and easily understandable system for provenance data generation usable by both security experts and other non-security researchers from areas like machine learning and data science.

\subsection{Automated Attack Systems}
\subsubsection{Xanthus}
In conjunction with Unicorn~\cite{unicorn}, Han et al highlight the lack of publicly available provenance graph datasets and emphasize the deficits of those datasets that are made available to researchers. They propose a push-button orchestrated system~\cite{xanthus} for recreating test and training data for provenance-based intrusion detection systems. Xanthus introduces the idea of a distributable virtual machine image with configurable jobs for automated recreation of attack scenarios, but this system requires the user to orchestrate the attack and to ensure that the VM and job scripts are correctly configured to mimic real-life scenarios. This requires extensive systems knowledge and makes the framework difficult to use for other researchers focused in data science and not security.

\section{System Design}

\textsc{Flurry} is designed to work as both independent software modules or a joint pipeline to support provenance graph learning from data creation and storage to examination. In the following sections, we detail each stage of this process which can work jointly or independently at the discretion of the user. Throughout the system, we offer methods for importing and exporting data in multiple formats for maximum adaptability.

\begin{figure} [ht]
  \includegraphics[width=\linewidth]{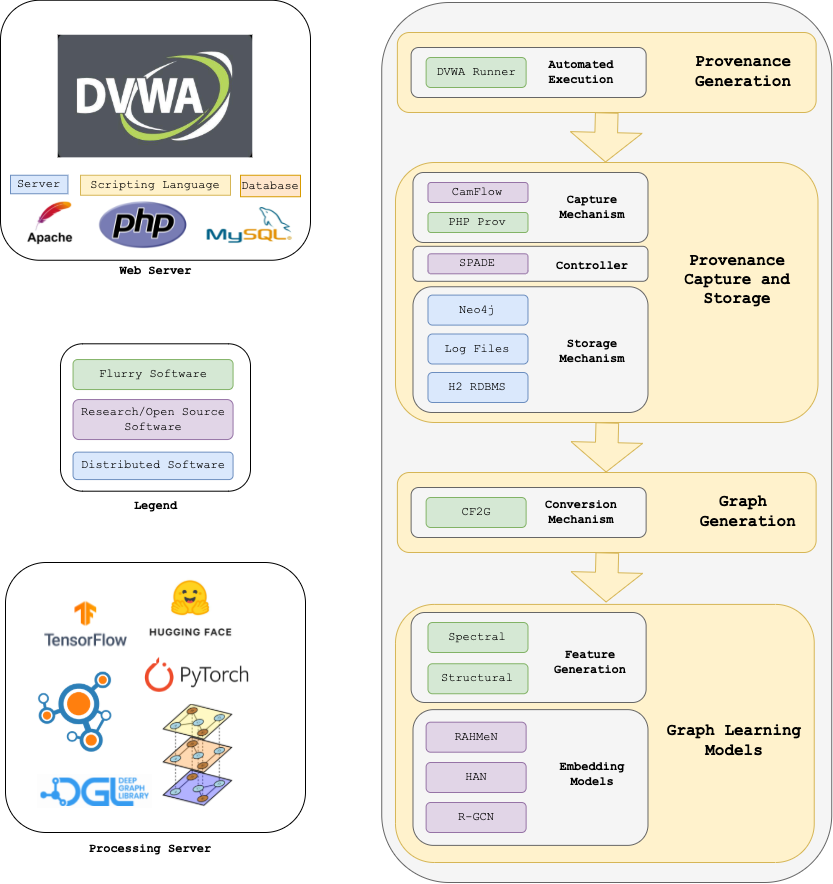}
  \caption{A system overview of the fast \textsc{Flurry} framework.}
  \label{fig:flurry}
\end{figure}

\subsection{Cyber Attack Automation}

For \textsc{Flurry} v1.0, we focus on generating provenance data for web-based cyber attacks. In this architecture, provenance may be captured for data exchanged across sockets, server-client or peer-to-peer requests and responses, software or other attachments queried for and returned, file reads and writes and memory access on both server and client, and running processes on the server and client, to name a few points of interest. In order to simulate web-based cyber attacks on a web server, we run an instance of the Damn Vulnerable Web Application~\cite{dvwa} on the Fedora virtual machine that is configured with a traditional XAMPP stack including Apache2 web server and MySQL database for storage. This PHP-based application that allows security students and professionals to practice skills and tools in a legal environment. Because it is vulnerable, it is strongly recommended not to run the application on any web server with real user information; therefore, we also recommend not uploading proprietary or private data to \textsc{Flurry} to prevent exfiltration.

In order to start the DVWA application and web services needed to run it, we provide a graphical user interface for researchers to start all dependencies with the push of a button. Once initiated, the user may select one of the attack scenarios to run and generate provenance, or a benign scenario which corresponds to one of the attacks from our \textbf{DVWA Runner} application. This level of abstraction over our automated scripts allows for rapid generation of gigabytes of data in mere seconds even by inexperienced users. More details about the attacks \textsc{Flurry} currently supports and the benign behavior is provided in the datasets section.

As provenance is generated on the backend, the user will see our web driver service run the scenario in real time on the screen. This allows for full transparency of the attack process so that researchers are aware of how their data was made. In addition to the user interface, we provide the Python scripts themselves so that more experienced users may engineer new scenarios or make adjustments according to their needs.

\subsection{Data Provenance Capture and Management}

The capture of data provenance is an entire subset field of data engineering. At the systems level, there are multiple layers, or planes, at which data is exchanged. In the \textit{kernel space}, systems like Hi-Fi~\cite{hifi}, the Linux Provenance Modules (LPM)~\cite{lpm} framework, and CamFlow~\cite{camflow} capture IPC mechanisms, network activity, and access to the kernel itself from the perspective of the operating system. These systems use hooks into security access mechanisms to write provenance to a buffer which is then typically translated into user space. In the \textit{user space}, we can divide provenance into even more fine-grained layers. For example, one may want to isolate application-layer provenance to particular files the user is modifying. Details of what was modified will not be recorded at the kernel level, but we may want to semantically tie kernel-level security access with data modification at the application level. This can be further extended to browser-level provenance if we are using a browser application to view or modify the file, and may even go to multi-host or network-level provenance if we are sharing this file through a file transfer service like FTP. Part of the data engineering required for good data provenance is determining the granularity at which to capture metadata and how to semantically tie these layers together. On the one hand, capturing too little or too coarse an image of the system execution leads to over-summarization and can cause analysts to miss anomalies. However, too much detail or association leads to \textit{dependency explosion}, or the phenomena of data becoming so entangled that the strongest semantic ties are lost.

In \textsc{Flurry}, we are proposing to support a multi-plane approach for provenance graphs for representation learning, which to our knowledge is a novel addition to this subfield. At the kernel level, we use \textbf{CamFlow}, which is a Linux security module, to capture provenance through kernel hooks. Data provenance which is captured by this service include security access to system information like inodes, security contexts, timestamps and jiffies, which processes accessed kernel objects, and more. The full description of CamFlow provenance node and edge types is well-recorded in their documentation~\cite{camflow}. This provenance information is published to a ring buffer which is then captured by the CamFlow daemon. In the default configuration of \textsc{Flurry}, we write this provenance to a log file on the virtual machine, but it may be additionally configured to use a unix pipe or a messaging service in order to convey this information dynamically. This allows the system to work in a non-stop manner without the need for static data files.

In addition to kernel-level provenance, \textsc{Flurry} adds hooks to PHP-based source code in DVWA in order to capture application-level provenance. Such provenance allows us to trace the flow of data within the web application such as where they came from, how they came to be in their present state, who or what acted upon them. Here in our application, events such as who inputs the data, how it is transformed to an HTTP request, and subsequent server responses are recorded. This additional layer may be combined with kernel access information to provide semantic context to what is happening at the kernel level. In order to capture both layers of provenance information in our audit logs, we  integrate the DVWA application source code with the Camflow Userspace API. For such integrations, we have also used a library called PHP-CPP, developed by Copernicus. This combination of application-level and kernel-level provenance gives a wider lens into system execution and a larger plane for anomaly detection.

Over time, storing provenance as log files is both space-consuming and unexpressive. Researchers may want to store provenance in a graph database as opposed to logs. This not only allows for provenance to be stored as nodes and edges, but also supports querying for specific graph structures from the large amount of data. For this reason, \textsc{Flurry} can also be configured to push data to \textbf{Neo4j}~\cite{neo4j}, a graph database which can be refined and queried. Storage can also be managed in the \textsc{Flurry} environment through the \textbf{SPADE}~\cite{spade} controller, an open-source provenance management software created by Gehani et al. SPADE offers different provenance capture methods including a CamFlow interface, analyzers, and transformers to perform execution partitioning and graph summarization. SPADE's purpose is also to serve as a control interface to port CamFlow data into Neo4j or even relational databases like H2. As a research software, some of these capabilities remain unimplemented or run with error. In future work, we plan to expand on some of these additional capabilities and contribute to their improvement.

\subsection{Provenance Graph Conversion}

Related works in provenance graph representation learning have written their own scripts for conversion of provenance data into complex graphs~\cite{sigl, unicorn, streamspot}. While this serves as a stop-gap solution, ultimately these works have not focused on reusability or extensibility of conversion. As opposed to scripts for fixed use cases, we add to the field a software tool which supports conversion of data from many input forms created by mainstream provenance capture tools into software or serialized graphs usable by machine learning frameworks. In \textsc{Flurry} v1.0, we call this \textbf{CF2G}, which stands for ``CamFlow to graph". CF2G accepts as input data serialized in the W3C-PROV~\cite{w3c-prov} JSON format, or the SPADE~\cite{spade} JSON format, both of which are published by the CamFlow daemon and configurable via the camflowd service. From the command line interface, CF2G parses the input graph and stores it as an internal graph. Using the CLI or the packaged library, a researcher may create JSON-exportable network representations of the stored provenance, call for the creation of a NetworkX~\cite{networkx} graph, or construct a Deep Graph Library~\cite{dgl} graph. CF2G also supports statistical output of the recorded provenance such as node types, edge types, and overall dimensionality. Currently, we support direct library calls or serialized and compressed export of the graphs to the next module.

\subsection{Feature Engineering}
Graph convolutional network models for learning node embeddings rely on iteratively aggregating a node's information with the information of its neighbors.
This process hinges on each node having an initial node feature vector which can be used in the aggregation process.
For this work we use structural node information including node degree and custom spectral methods for heterogeneous graphs extended from~\cite{eigenmaps}, which performs an eigendecomposition using the LOBPCG~\cite{lobpcg2002, lobpcg2018} estimator on a block matrix~\cite{multilayer_bian} built around the normalized Laplacian matrices of each relation type in the graph.

\subsection{GNN Trainer}
In addition to our system provenance collection and graph generation tools, we provide a unified graph neural network trainer based on DGL and PyTorch, inspired by the Trainer class provided by HuggingFace for training Transformer models on NLP tasks. Our training framework consists of a base Trainer class that defines the core training and evaluation functions for graph classification using GNN models, and a TrainingArguments class that defines training and evaluation hyperparameters and additional behavior. We provide a set of default callback functions that define normal training operations, including logging, saving model checkpoints, and early stopping using validation metrics. Our framework also supports custom user-defined callback functions to further customize the Trainer's behavior during model training and evaluation. Our Trainer framework utilizes DGL's graph batching and message-passing functionality to provide a standardized yet flexible framework for training graph neural models. We support standard PyTorch loss functions, such as cross-entropy loss, and also custom-defined trainable loss functions that can be used for unsupervised training on graphs. Our framework currently supports graph classification tasks, and we plan to extend our Trainer framework to support node classification and edge prediction tasks in the future.

\section{Threat Models} \label{sec:datasets}
The exact datasets we used for experiments will be publicly available. Table~\ref{table:datasets} shows the size of the graphs in terms of nodes and edges. Relation types are defined as unique source node, edge type, destination node three-tuples. The execution involves navigating to the tab on the DVWA website that corresponds to the attack and performing some attack or benign behavior. The attacks covered in this dataset and by the current version of \textsc{Flurry} cover more than 50\% of the most popular attack vectors cyber criminals choose for web-based applications~\cite{positivetechnologies2018report}.

\begin{center}
\begin{table*}
\begin{tabular}{|c | c | c | c | c |}
\hline
 \textbf{Attack Vector} & \textbf{Class Type} & \textbf{Avg. \# of Nodes} & \textbf{Avg. \# of Edges} & \textbf{Avg. \# of Relation Types}\\
 \hline
 XSS Stored & Benign & 19,436 & 463,682 & 30 \\
 & Attack & 19,547 & 480,179 & 31 \\
 \hline
 XSS Reflected & Benign & 33,481 & 824,276 & 31 \\
 & Attack & 24,435 & 666,705 & 32 \\
 \hline
 XSS DOM & Benign & 31,243 & 753,519 & 30 \\
 & Attack & 30,261 & 751,133 & 32 \\
 \hline
 Command Line Injection & Benign & 32,996 & 793,318 & 29 \\
 & Attack & 25,720 & 631,633 & 30 \\
 \hline
 SQL Injection & Benign & 22,903 & 543,266 & 30 \\
 & Attack & 29,576 & 733,375 & 30 \\
 \hline
 Brute Force & Benign & 21,518 & 517,101 & 30 \\
 & Attack & 418 & 416 & 1 \\
 \hline
\end{tabular}
\caption{Description of Attack Graph Datasets}
\label{table:datasets}
\end{table*}
\end{center}

\subsection*{Cross-Site Scripting} Cross-site scripting allows an attacker to inject JavaScript in a web application in order to modify the web page viewed by the user or perform some other malicious action such as data exfiltration or modification. Like other injection attacks, the vulnerability usually exists because of a lack of data input sanitization from web forms or requests.

\textbf{XSS Reflected.} Reflected scripts are non-persistent, meaning that they are ephemerally injected into the HTTP client request processed by the web server. If the web server does not properly read the HTML escape sequences when reading the request, the JavaScript in the request will be executed on the page on the server side. In the \textsc{Flurry} dataset, we create a benign execution by filling out and submitting the web form question (``What is your name?"). For the attack, we inject JavaScript in the web form to pop up a text box and corresponding message after submission.

\textbf{XSS Stored.} A stored scripted attack is not sent by the client, but rather stored in the server database and thus is persistent as it can be re-executed. For example, an attacker may post a message to a forum with an embedded script. When other clients go to read the message, the attacker's script will be executed when the message is read/returned~\cite{positivetechnologiesxss}. In \textsc{Flurry}, we automate posting a non-malicious message to a forum on the XSS Stored tab for benign execution. For the attack, we inject a script in the message which creates a bot-style malicious post every time the tab is visited.

\textbf{XSS DOM.} A DOM-style attack can be modified on the client side and thus does not require access to the web server or poor sanitization on the server's part. It relies on poor reading of HTML escape sequences in the client's document object model. The XSS DOM tab in the Damn Vulnerable Web Application asks the user to choose a language, and this setting is then part of the DOM and shown in the public-facing URL. In \textsc{Flurry}, we modify the DOM by injecting the script into the web URL. Our script creates a pop-up similar to the XSS Reflected attack. For benign behavior, we automate selecting English as the default language.

\subsection*{Injection}
While cross-site scripting is a style of injection attack specific to JavaScript and the web page, other types of injections such as SQL for database access/modification and command line injection can be used to cause remote code execution, privilege escalation, data exfiltration, and other types of more consequential attacks.

\textbf{Command Line Injection.} When web submissions are not properly sanitized, system commands may be included and will be executed on the server machine. In DVWA, the Command Line Injection tab asks the user to provide an IP address to ping. For the benign execution, we provide localhost 127.0.0.1 for pinging. For the attack, we inject \texttt{pwd} to print the current working directory where the server is executing the ping.

\textbf{SQL Injection.} In the DVWA SQL Injection tab, the user is prompted for an ID. In a benign execution, this ID is included as part of a SQL query which is looked up in the internal database. In the attack automation, we include a semicolon and additional SQL query for other user IDs.

\subsection*{Password Cracking}
In addition to injection-style attacks, we wanted to expand into password crackers, which make up a significant portion of web application-based cyber attacks~\cite{positivetechnologies2018report}. In future versions of \textsc{Flurry}, we may further look into more advanced rainbow table or dictionary style password attacks.

\textbf{Brute Force.} There are an array of automated tools for brute force; we use the Hydra~\cite{hydra} password cracking tool. We provide an attack script which executes the hydra command and provide a dictionary with the 100 most commonly used passwords to try. For benign execution, we automate logging in with the correct credentials.

\section{Experiments}

\begin{table*}[hbt!]
%\begin{tabular}{|l|lll|lll|lll|}
\begin{tabular}{|l|ccc|ccc|ccc|}
\hline
              &                & HAN             &                 &               & R-GCN         &               &                 & PROV-GEm         &           \\ \cline{2-10}
              & F1             & Precision       & Recall         & F1            & Precision     & Recall        & F1              & Precision       & Recall    \\ \hline
brute-force   & 99.51$\pm$0.97 & 100.0$\pm$0.0 & 99.35$\pm$1.29 & 100.0$\pm$0.0 & 100.0$\pm$0.0 & 100.0$\pm$0.0 & 100.0$\pm$0.0 & 100.0$\pm$0.0 & 100.0$\pm$0.0 \\
cl-injection  & 63.35$\pm$8.07 & 81.44$\pm$15.97 & 65.57$\pm$12.98 & 70.60$\pm$8.26 & 69.16$\pm$23.59 & 79.84$\pm$12.95 & 58.08$\pm$9.56 & 63.72$\pm$15.37 & 60.00$\pm$12.76 \\
sql-injection & 71.32$\pm$17.85 & 71.43$\pm$16.39 & 83.30$\pm$17.35 & 81.66$\pm$9.19 & 73.07$\pm$15.60 & 89.26$\pm$3.39 & 71.30$\pm$10.61 & 84.59$\pm$5.55 & 69.41$\pm$10.42 \\
xss-dom       & 91.23$\pm$4.33 & 84.83$\pm$8.43 & 97.50$\pm$5.00 & 80.70$\pm$9.97 & 81.66$\pm$16.51 & 82.92$\pm$11.95 & 91.23$\pm$4.33 & 84.83$\pm$8.43 & 97.50$\pm$5.00 \\
xss-reflected & 90.35$\pm$4.84 & 88.79$\pm$9.83 & 93.63$\pm$4.05 & 84.22$\pm$4.55 & 85.93$\pm$8.53 & 83.64$\pm$6.37 & 88.87$\pm$7.23 & 83.63$\pm$14.22 & 95.16$\pm$4.78 \\
xss-stored    & 52.80$\pm$8.36 & 71.87$\pm$14.48 & 53.96$\pm$10.51 & 48.39$\pm$6.64 & 71.00$\pm$23.63 & 51.59$\pm$6.52 & 53.50$\pm$8.48 & 64.59$\pm$15.92 & 55.93$\pm$12.64 \\ \hline
\end{tabular}
\caption{\label{tab:clf-results}Graph classification results using generated \textsc{Flurry} data for six system attack types.}
\end{table*}

In this section, we describe the anomaly detection experiments we perform on system provenance data collected by \textsc{Flurry}. For each of the six attack types characterized in Section~\ref{sec:datasets}, we simulate 100 benign operations and 100 attack operations. We convert the system provenance data into heterogeneous graphs and partition the datasets into five cross validation folds to ensure consistent and reproducible results. For each attack type, we conduct graph classification experiments using our Trainer class with three different heterogeneous graph embedding frameworks from graph representational learning literature---R-GCN~\cite{rgcn}, HAN~\cite{han}, and PROV-GEm~\cite{prov-gem}---implemented in DGL. In each experimental setup, one cross validation fold is held out as a test set, and the graph embedding models are trained with the remaining provenance graphs using cross entropy loss. Our results on the graph anomaly detection task are presented in Table~\ref{tab:clf-results}. We report the mean and standard deviation over five-fold cross validation of precision, recall, and F1 score. The code for our experiments along with our graph neural network training framework are provided at\footnote{https://github.com/NASCL/dgl-trainer.git}

Due to the nature of brute force attacks, their system provenance generates graphs containing only a single relation type: (socket, WasDerivedFrom, socket). As such, these graphs are trivially easy to distinguish from benign system behavior, and all three of the graph neural network models were able to perform perfectly on the generated data. By contrast, for the two injection-style attacks---command line injection and SQL injection---anomalous behavior is much more difficult to identify from benign provenance graphs generate during normal application use. Both scenarios exhibited high variance in performance across cross validation folds, suggesting that both injection-style attacks can vary in their graph properties.

In our experiments, we found that the SQL injection and command line injection attacks were more difficult to detect than others from the datasets that were gathered. Considering the system execution, the kernel-level provenance and database-level provenance for both benign and attack scenarios are similar as both are ultimately SQL queries or file system reads/writes. It is evident from our experiments that single-layer provenance is not able to capture anomalies for all attack types; some attacks are more evident at different layers of the system execution. Currently, existing provenance capture tools have provided some anomaly detection capability, but there is a lack of tools for attack types which are not evident at the kernel level. Our results here are consistent with the postulation that multi-layer provenance would be more effective.

A similar phenomenon can be noted with our results in the stored cross-site scripting scenario. The attack behavior is a bot-style message. The attack itself is not much different from benign posting behavior, but the frequency of the message may be noted as anomalous in a graph learning system which models time-series information.

Our models performed best on the cross-site reflected and DOM scenarios. For the reflected threat model, the attack system execution behavior is distinctly different as it pulls up a pop-up window which does not exist in the benign version. Similarly, the DOM model also generates a pop-up window that does not exist in the benign scenario. Both of these generate unique provenance graph substructures which are detectable by the models as anomalous in the provenance layers we examined. A significant finding of our experiments is that whole-scope anomaly detection in provenance graphs will continue to require constant vigilance in multiple layers of the target system - no single layer solution will be sufficient to catch even a narrow variety of attacks.

\section{Discussion and Future Work}

The \textsc{Flurry} system is a novel, initial contribution to automated provenance graph generation and machine learning and thus has room for testing and improvement. The system currently writes logs to files intermittently. This design is intentional as we do not provide graph summarization beyond the de-duplication that CamFlow~\cite{camflow} supports, and thus would overwhelm the downstream applications with log data if it were piped directly. Future versions will include a graph summarization element and the ability to work with streaming data as well as static graphs.

One aspect of browser-level provenance which we have not yet implemented in \textsc{Flurry} is to capture and compare user entries into forms. We propose in future work to develop a browser provenance extension to \textsc{Flurry} which will scrape these forms and use similarity metrics of routine entries as a potential learning feature. As a practical example, an injected SQL query will be quite dissimilar to an expected web form entry for a username. Locality sensitive hashing could provide both a distance measure that could improve model results for this attack type and obfuscation for user privacy. This experimental result and potential solution reveal the importance of multi-layer provenance, as different attacks will be more evident at different layers of the system.

In future versions of \textsc{Flurry}, we intend to design a streaming graph approach to the pipeline in addition to segmented graphs in order to better capture attack rates. This approach has been explored in provenance graph learning systems~\cite{unicorn, streamspot} and would be a useful addition to our automated framework. It gives an additional insight to data provenance that is temporal as opposed to solely spatial. This would also likely improve results in other frequency-based vectors like flooding in denial-of-service or brute-force attacks.

\section{Conclusion}
\textsc{Flurry} advances the state of the art by extending both ways in the process to create a full data pipeline for provenance graph representation learning. First, we expand the user-friendliness and quality assurance of the traces by providing a GUI for ease of interaction with the attack scenarios. We also use a web driver to automate the attacks so that the researcher creating the data can watch the attack as it happens. The data is then captured by provenance tools pre-installed on our distributable VM image. While highly configurable, these tools are also ready at deployment to capture provenance at the click of a button. We provide additional tools to export the raw data or process it into provenance graphs. For maximum flexibility we provide a library tool for converting the raw logs into NetworkX or Deep Graph Library graphs, exportable JSON, or compressed and serialized formats which can be manually uploaded into other machine learning frameworks. Finally, we provide a collection of deep learning models for examination of these provenance graphs as they are generated. Our experiments and results reveal the capability of \textsc{Flurry} as a provenance graph generation system from cyber attack to anomaly detection. We offer this system as a tool for the research community to continue to advance the state of the art in provenance graph representation learning.

%\section{Acknowledgments}

%\section{Appendices}
%\input{appendix.tex}
%\newpage
\bibliographystyle{ACM-Reference-Format}
\bibliography{master}


\begin{thebibliography}{33}

%%% the Web address.

\ifx \showCODEN    \undefined \def \showCODEN     #1{\unskip}     \fi
\ifx \showDOI      \undefined \def \showDOI       #1{#1}\fi
\ifx \showISBNx    \undefined \def \showISBNx     #1{\unskip}     \fi
\ifx \showISBNxiii \undefined \def \showISBNxiii  #1{\unskip}     \fi
\ifx \showISSN     \undefined \def \showISSN      #1{\unskip}     \fi
\ifx \showLCCN     \undefined \def \showLCCN      #1{\unskip}     \fi
\ifx \shownote     \undefined \def \shownote      #1{#1}          \fi
\ifx \showarticletitle \undefined \def \showarticletitle #1{#1}   \fi
\ifx \showURL      \undefined \def \showURL       {\relax}        \fi
% invisible to TeX
\providecommand\bibfield[2]{#2}
\providecommand\bibinfo[2]{#2}
\providecommand\natexlab[1]{#1}
\providecommand\showeprint[2][]{arXiv:#2}

\bibitem[\protect\citeauthoryear{Bagavathi and Krishnan}{Bagavathi and
  Krishnan}{2019}]%
        {multinet}
\bibfield{author}{\bibinfo{person}{Arunkumar Bagavathi} {and}
  \bibinfo{person}{Siddharth Krishnan}.} \bibinfo{year}{2019}\natexlab{}.
\newblock \showarticletitle{Multi-Net: A Scalable Multiplex Network Embedding
  Framework}. In \bibinfo{booktitle}{\emph{Complex Networks and Their
  Applications VII}}, \bibfield{editor}{\bibinfo{person}{Luca~Maria Aiello},
  \bibinfo{person}{Chantal Cherifi}, \bibinfo{person}{Hocine Cherifi},
  \bibinfo{person}{Renaud Lambiotte}, \bibinfo{person}{Pietro Li{\'o}}, {and}
  \bibinfo{person}{Luis~M. Rocha}} (Eds.). \bibinfo{publisher}{Springer
  International Publishing}, \bibinfo{address}{Cham},
  \bibinfo{pages}{119--131}.
\newblock
\showISBNx{978-3-030-05414-4}


\bibitem[\protect\citeauthoryear{Bates, Tian, Butler, and Moyer}{Bates
  et~al\mbox{.}}{2015}]%
        {lpm}
\bibfield{author}{\bibinfo{person}{Adam Bates}, \bibinfo{person}{Dave Tian},
  \bibinfo{person}{Kevin R.~B. Butler}, {and} \bibinfo{person}{Thomas Moyer}.}
  \bibinfo{year}{2015}\natexlab{}.
\newblock \showarticletitle{Trustworthy Whole-System Provenance for the Linux
  Kernel}. In \bibinfo{booktitle}{\emph{Proceedings of the 24th USENIX
  Conference on Security Symposium}} (Washington, D.C.)
  \emph{(\bibinfo{series}{SEC'15})}. \bibinfo{publisher}{USENIX Association},
  \bibinfo{address}{USA}, \bibinfo{pages}{319–334}.
\newblock
\showISBNx{9781931971232}


\bibitem[\protect\citeauthoryear{Belkin and Niyogi}{Belkin and Niyogi}{2003}]%
        {eigenmaps}
\bibfield{author}{\bibinfo{person}{Mikhail Belkin} {and}
  \bibinfo{person}{Partha Niyogi}.} \bibinfo{year}{2003}\natexlab{}.
\newblock \showarticletitle{Laplacian eigenmaps for dimensionality reduction
  and data representation}.
\newblock \bibinfo{journal}{\emph{Neural computation}} \bibinfo{volume}{15},
  \bibinfo{number}{6} (\bibinfo{year}{2003}), \bibinfo{pages}{1373--1396}.
\newblock


\bibitem[\protect\citeauthoryear{Bianconi}{Bianconi}{2018}]%
        {multilayer_bian}
\bibfield{author}{\bibinfo{person}{Ginestra Bianconi}.}
  \bibinfo{year}{2018}\natexlab{}.
\newblock \bibinfo{booktitle}{\emph{Multilayer Networks: Structure and
  Function}}.
\newblock \bibinfo{publisher}{Oxford university press}.
\newblock


\bibitem[\protect\citeauthoryear{digininja}{digininja}{2020}]%
        {dvwa}
\bibfield{author}{\bibinfo{person}{digininja}.}
  \bibinfo{year}{2020}\natexlab{}.
\newblock \bibinfo{title}{Damn Vulnerable Web Application}.
\newblock \bibinfo{howpublished}{\url{https://github.com/digininja/DVWA}}.
\newblock


\bibitem[\protect\citeauthoryear{Dong, Chawla, and Swami}{Dong
  et~al\mbox{.}}{2017}]%
        {metapath2vec}
\bibfield{author}{\bibinfo{person}{Yuxiao Dong}, \bibinfo{person}{Nitesh~V
  Chawla}, {and} \bibinfo{person}{Ananthram Swami}.}
  \bibinfo{year}{2017}\natexlab{}.
\newblock \showarticletitle{metapath2vec: Scalable representation learning for
  heterogeneous networks}. In \bibinfo{booktitle}{\emph{SIGKDD}}.
  \bibinfo{pages}{135--144}.
\newblock


\bibitem[\protect\citeauthoryear{Duersch, Shao, Yang, and Gu}{Duersch
  et~al\mbox{.}}{2018}]%
        {lobpcg2018}
\bibfield{author}{\bibinfo{person}{Jed~A Duersch}, \bibinfo{person}{Meiyue
  Shao}, \bibinfo{person}{Chao Yang}, {and} \bibinfo{person}{Ming Gu}.}
  \bibinfo{year}{2018}\natexlab{}.
\newblock \showarticletitle{A robust and efficient implementation of {LOBPCG}}.
\newblock \bibinfo{journal}{\emph{SIAM Journal on Scientific Computing}}
  \bibinfo{volume}{40}, \bibinfo{number}{5} (\bibinfo{year}{2018}),
  \bibinfo{pages}{C655--C676}.
\newblock


\bibitem[\protect\citeauthoryear{Gehani and Tariq}{Gehani and Tariq}{2012}]%
        {spade}
\bibfield{author}{\bibinfo{person}{Ashish Gehani} {and} \bibinfo{person}{Dawood
  Tariq}.} \bibinfo{year}{2012}\natexlab{}.
\newblock \showarticletitle{{{SPADE}}: {{Support}} for Provenance Auditing in
  Distributed Environments}.
\newblock In \bibinfo{booktitle}{\emph{Middleware 2012}},
  \bibfield{editor}{\bibinfo{person}{Priya Narasimhan} {and}
  \bibinfo{person}{Peter Triantafillou}} (Eds.). \bibinfo{series}{Lecture Notes
  in Computer Science}, Vol.~\bibinfo{volume}{7662}.
  \bibinfo{publisher}{{Springer Berlin Heidelberg}}, \bibinfo{pages}{101--120}.
\newblock
\showISBNx{978-3-642-35169-3}
\urldef\tempurl%
\url{https://doi.org/10.1007/978-3-642-35170-9_6}
\showDOI{\tempurl}


\bibitem[\protect\citeauthoryear{Grover and Leskovec}{Grover and
  Leskovec}{2016}]%
        {node2vec}
\bibfield{author}{\bibinfo{person}{Aditya Grover} {and} \bibinfo{person}{Jure
  Leskovec}.} \bibinfo{year}{2016}\natexlab{}.
\newblock \showarticletitle{node2vec: Scalable feature learning for networks}.
  In \bibinfo{booktitle}{\emph{Proceedings of the 22nd ACM SIGKDD international
  conference on Knowledge discovery and data mining}}.
  \bibinfo{pages}{855--864}.
\newblock


\bibitem[\protect\citeauthoryear{Hamilton, Ying, and Leskovec}{Hamilton
  et~al\mbox{.}}{2017}]%
        {graphsage}
\bibfield{author}{\bibinfo{person}{Will Hamilton}, \bibinfo{person}{Zhitao
  Ying}, {and} \bibinfo{person}{Jure Leskovec}.}
  \bibinfo{year}{2017}\natexlab{}.
\newblock \showarticletitle{Inductive representation learning on large graphs}.
  In \bibinfo{booktitle}{\emph{NeurIPS}}. \bibinfo{pages}{1024--1034}.
\newblock


\bibitem[\protect\citeauthoryear{Han}{Han}{2018}]%
        {streamspot-dataset}
\bibfield{author}{\bibinfo{person}{Xueyuan Han}.}
  \bibinfo{year}{2018}\natexlab{}.
\newblock \bibinfo{title}{{StreamSpot Dataset}}.
\newblock
\newblock
\urldef\tempurl%
\url{https://doi.org/10.7910/DVN/83KYJY}
\showDOI{\tempurl}


\bibitem[\protect\citeauthoryear{Han, Mickens, Gehani, Seltzer, and
  Pasquier}{Han et~al\mbox{.}}{2020a}]%
        {xanthus}
\bibfield{author}{\bibinfo{person}{Xueyuan Han}, \bibinfo{person}{James
  Mickens}, \bibinfo{person}{Ashish Gehani}, \bibinfo{person}{Margo~I.
  Seltzer}, {and} \bibinfo{person}{Thomas F.~J.{-}M. Pasquier}.}
  \bibinfo{year}{2020}\natexlab{a}.
\newblock \showarticletitle{Xanthus: Push-button Orchestration of Host
  Provenance Data Collection}.
\newblock \bibinfo{journal}{\emph{CoRR}}  \bibinfo{volume}{abs/2005.04717}
  (\bibinfo{year}{2020}).
\newblock
\showeprint[arXiv]{2005.04717}
\urldef\tempurl%
\url{https://arxiv.org/abs/2005.04717}
\showURL{%
\tempurl}


\bibitem[\protect\citeauthoryear{Han, Pasquier, Bates, Mickens, and
  Seltzer}{Han et~al\mbox{.}}{2020b}]%
        {unicorn}
\bibfield{author}{\bibinfo{person}{Xueyuan Han}, \bibinfo{person}{Thomas
  F.~J.{-}M. Pasquier}, \bibinfo{person}{Adam Bates}, \bibinfo{person}{James
  Mickens}, {and} \bibinfo{person}{Margo~I. Seltzer}.}
  \bibinfo{year}{2020}\natexlab{b}.
\newblock \showarticletitle{{UNICORN:} Runtime Provenance-Based Detector for
  Advanced Persistent Threats}.
\newblock \bibinfo{journal}{\emph{CoRR}}  \bibinfo{volume}{abs/2001.01525}
  (\bibinfo{year}{2020}).
\newblock
\showeprint[arxiv]{2001.01525}
\urldef\tempurl%
\url{http://arxiv.org/abs/2001.01525}
\showURL{%
\tempurl}


\bibitem[\protect\citeauthoryear{Han, Yu, Pasquier, Li, Rhee, Mickens, Seltzer,
  and Chen}{Han et~al\mbox{.}}{2021}]%
        {sigl}
\bibfield{author}{\bibinfo{person}{Xueyuan Han}, \bibinfo{person}{Xiao Yu},
  \bibinfo{person}{Thomas Pasquier}, \bibinfo{person}{Ding Li},
  \bibinfo{person}{Junghwan Rhee}, \bibinfo{person}{James Mickens},
  \bibinfo{person}{Margo Seltzer}, {and} \bibinfo{person}{Haifeng Chen}.}
  \bibinfo{year}{2021}\natexlab{}.
\newblock \bibinfo{title}{SIGL: Securing Software Installations Through Deep
  Graph Learning}.
\newblock
\newblock
\showeprint[arxiv]{2008.11533}~[cs.CR]


\bibitem[\protect\citeauthoryear{Kapoor, Melton, Ridenhour, Krishnan, and
  Moyer}{Kapoor et~al\mbox{.}}{2021}]%
        {prov-gem}
\bibfield{author}{\bibinfo{person}{Maya Kapoor}, \bibinfo{person}{Joshua
  Melton}, \bibinfo{person}{Michael Ridenhour}, \bibinfo{person}{Siddharth
  Krishnan}, {and} \bibinfo{person}{Thomas Moyer}.}
  \bibinfo{year}{2021}\natexlab{}.
\newblock \showarticletitle{PROV-GEM: Automated Provenance Analysis Framework
  using Graph Embeddings}. In \bibinfo{booktitle}{\emph{2021 20th IEEE
  International Conference on Machine Learning and Applications (ICMLA)}}.
  \bibinfo{pages}{1720--1727}.
\newblock
\urldef\tempurl%
\url{https://doi.org/10.1109/ICMLA52953.2021.00273}
\showDOI{\tempurl}


\bibitem[\protect\citeauthoryear{Kipf and Welling}{Kipf and Welling}{2017}]%
        {kipf_gcn}
\bibfield{author}{\bibinfo{person}{Thomas~N. Kipf} {and} \bibinfo{person}{Max
  Welling}.} \bibinfo{year}{2017}\natexlab{}.
\newblock \showarticletitle{Semi-Supervised Classification with Graph
  Convolutional Networks}. In \bibinfo{booktitle}{\emph{ICLR}}.
\newblock


\bibitem[\protect\citeauthoryear{Liu, Huang, Yu, Fan, and Dong}{Liu
  et~al\mbox{.}}{2020}]%
        {fame}
\bibfield{author}{\bibinfo{person}{Zhijun Liu}, \bibinfo{person}{Chao Huang},
  \bibinfo{person}{Yanwei Yu}, \bibinfo{person}{Baode Fan}, {and}
  \bibinfo{person}{Junyu Dong}.} \bibinfo{year}{2020}\natexlab{}.
\newblock \showarticletitle{Fast Attributed Multiplex Heterogeneous Network
  Embedding}. In \bibinfo{booktitle}{\emph{Proceedings of the 29th ACM
  International Conference on Information \& Knowledge Management}}.
  \bibinfo{pages}{995--1004}.
\newblock


\bibitem[\protect\citeauthoryear{Manzoor, Momeni, Venkatakrishnan, and
  Akoglu}{Manzoor et~al\mbox{.}}{2016}]%
        {streamspot}
\bibfield{author}{\bibinfo{person}{Emaad~A. Manzoor}, \bibinfo{person}{Sadegh
  Momeni}, \bibinfo{person}{Venkat~N. Venkatakrishnan}, {and}
  \bibinfo{person}{Leman Akoglu}.} \bibinfo{year}{2016}\natexlab{}.
\newblock \bibinfo{title}{Fast Memory-efficient Anomaly Detection in Streaming
  Heterogeneous Graphs}.
\newblock
\newblock
\showeprint[arxiv]{1602.04844}~[cs.SI]


\bibitem[\protect\citeauthoryear{Neo4j}{Neo4j}{2012}]%
        {neo4j}
\bibfield{author}{\bibinfo{person}{Neo4j}.} \bibinfo{year}{2012}\natexlab{}.
\newblock \bibinfo{title}{Neo4j - The World’s Leading Graph Database}.
\newblock
\newblock
\urldef\tempurl%
\url{http://neo4j.org/}
\showURL{%
\tempurl}


\bibitem[\protect\citeauthoryear{{NetworkX developer team}}{{NetworkX developer
  team}}{2014}]%
        {networkx}
\bibfield{author}{\bibinfo{person}{{NetworkX developer team}}.}
  \bibinfo{year}{2014}\natexlab{}.
\newblock \bibinfo{title}{NetworkX}.
\newblock
\newblock
\urldef\tempurl%
\url{https://networkx.github.io/}
\showURL{%
\tempurl}


\bibitem[\protect\citeauthoryear{Pasquier, Han, Goldstein, Moyer, Eyers,
  Seltzer, and Bacon}{Pasquier et~al\mbox{.}}{2017}]%
        {camflow}
\bibfield{author}{\bibinfo{person}{Thomas F.~J.{-}M. Pasquier},
  \bibinfo{person}{Xueyuan Han}, \bibinfo{person}{Mark Goldstein},
  \bibinfo{person}{Thomas Moyer}, \bibinfo{person}{David~M. Eyers},
  \bibinfo{person}{Margo~I. Seltzer}, {and} \bibinfo{person}{Jean Bacon}.}
  \bibinfo{year}{2017}\natexlab{}.
\newblock \showarticletitle{Practical Whole-System Provenance Capture}.
\newblock \bibinfo{journal}{\emph{CoRR}}  \bibinfo{volume}{abs/1711.05296}
  (\bibinfo{year}{2017}).
\newblock
\showeprint[arxiv]{1711.05296}
\urldef\tempurl%
\url{http://arxiv.org/abs/1711.05296}
\showURL{%
\tempurl}


\bibitem[\protect\citeauthoryear{Perozzi, Al-Rfou, and Skiena}{Perozzi
  et~al\mbox{.}}{2014}]%
        {deepwalk}
\bibfield{author}{\bibinfo{person}{Bryan Perozzi}, \bibinfo{person}{Rami
  Al-Rfou}, {and} \bibinfo{person}{Steven Skiena}.}
  \bibinfo{year}{2014}\natexlab{}.
\newblock \showarticletitle{Deepwalk: Online learning of social
  representations}. In \bibinfo{booktitle}{\emph{Proceedings of the 20th ACM
  SIGKDD international conference on Knowledge discovery and data mining}}.
  \bibinfo{pages}{701--710}.
\newblock


\bibitem[\protect\citeauthoryear{Pohly, McLaughlin, McDaniel, and Butler}{Pohly
  et~al\mbox{.}}{2012}]%
        {hifi}
\bibfield{author}{\bibinfo{person}{Devin~J. Pohly}, \bibinfo{person}{Stephen
  McLaughlin}, \bibinfo{person}{Patrick McDaniel}, {and} \bibinfo{person}{Kevin
  Butler}.} \bibinfo{year}{2012}\natexlab{}.
\newblock \showarticletitle{Hi-Fi: Collecting High-Fidelity Whole-System
  Provenance}. In \bibinfo{booktitle}{\emph{Proceedings of the 28th Annual
  Computer Security Applications Conference}} (Orlando, Florida, USA)
  \emph{(\bibinfo{series}{ACSAC '12})}. \bibinfo{publisher}{Association for
  Computing Machinery}, \bibinfo{address}{New York, NY, USA},
  \bibinfo{pages}{259–268}.
\newblock
\showISBNx{9781450313124}
\urldef\tempurl%
\url{https://doi.org/10.1145/2420950.2420989}
\showDOI{\tempurl}


\bibitem[\protect\citeauthoryear{{Positive Technologies}}{{Positive
  Technologies}}{2019}]%
        {positivetechnologies2018report}
\bibfield{author}{\bibinfo{person}{{Positive Technologies}}.}
  \bibinfo{year}{2019}\natexlab{}.
\newblock \bibinfo{title}{{Attacks on web applications: 2018 in review}}.
\newblock
\newblock
\urldef\tempurl%
\url{https://www.ptsecurity.com/ww-en/analytics/web-application-attacks-2019/}
\showURL{%
\tempurl}


\bibitem[\protect\citeauthoryear{{Positive Technologies}}{{Positive
  Technologies}}{2020}]%
        {positivetechnologiesxss}
\bibfield{author}{\bibinfo{person}{{Positive Technologies}}.}
  \bibinfo{year}{2020}\natexlab{}.
\newblock \bibinfo{title}{{What is a cross-site scripting (XSS) attack?}}
\newblock
\newblock
\urldef\tempurl%
\url{https://www.ptsecurity.com/ww-en/analytics/knowledge-base/what-is-a-cross-site-scripting-xss-attack/#5}
\showURL{%
\tempurl}


\bibitem[\protect\citeauthoryear{Schlichtkrull, Kipf, Bloem, van~den Berg,
  Titov, and Welling}{Schlichtkrull et~al\mbox{.}}{2018}]%
        {rgcn}
\bibfield{author}{\bibinfo{person}{Michael Schlichtkrull},
  \bibinfo{person}{Thomas~N. Kipf}, \bibinfo{person}{Peter Bloem},
  \bibinfo{person}{Rianne van~den Berg}, \bibinfo{person}{Ivan Titov}, {and}
  \bibinfo{person}{Max Welling}.} \bibinfo{year}{2018}\natexlab{}.
\newblock \showarticletitle{Modeling Relational Data with Graph Convolutional
  Networks}. In \bibinfo{booktitle}{\emph{The Semantic Web}}.
  \bibinfo{publisher}{Springer International Publishing},
  \bibinfo{pages}{593--607}.
\newblock
\showISBNx{978-3-319-93417-4}


\bibitem[\protect\citeauthoryear{Stathopoulos and Wu}{Stathopoulos and
  Wu}{2002}]%
        {lobpcg2002}
\bibfield{author}{\bibinfo{person}{Andreas Stathopoulos} {and}
  \bibinfo{person}{Kesheng Wu}.} \bibinfo{year}{2002}\natexlab{}.
\newblock \showarticletitle{A block orthogonalization procedure with constant
  synchronization requirements}.
\newblock \bibinfo{journal}{\emph{SIAM Journal on Scientific Computing}}
  \bibinfo{volume}{23}, \bibinfo{number}{6} (\bibinfo{year}{2002}),
  \bibinfo{pages}{2165--2182}.
\newblock


\bibitem[\protect\citeauthoryear{Tools}{Tools}{2021}]%
        {hydra}
\bibfield{author}{\bibinfo{person}{Kali~Linux Tools}.}
  \bibinfo{year}{2021}\natexlab{}.
\newblock \bibinfo{title}{Hydra}.
\newblock
\newblock
\urldef\tempurl%
\url{https://www.kali.org/tools/hydra/}
\showURL{%
\tempurl}


\bibitem[\protect\citeauthoryear{Veličković, Cucurull, Casanova, Romero,
  Liò, and Bengio}{Veličković et~al\mbox{.}}{2018}]%
        {gat}
\bibfield{author}{\bibinfo{person}{Petar Veličković},
  \bibinfo{person}{Guillem Cucurull}, \bibinfo{person}{Arantxa Casanova},
  \bibinfo{person}{Adriana Romero}, \bibinfo{person}{Pietro Liò}, {and}
  \bibinfo{person}{Yoshua Bengio}.} \bibinfo{year}{2018}\natexlab{}.
\newblock \showarticletitle{Graph {Attention} {Networks}}. In
  \bibinfo{booktitle}{\emph{ICLR}}.
\newblock


\bibitem[\protect\citeauthoryear{{W3C-PROV Working Group}}{{W3C-PROV Working
  Group}}{2013}]%
        {w3c-prov}
\bibfield{author}{\bibinfo{person}{{W3C-PROV Working Group}}.}
  \bibinfo{year}{2013}\natexlab{}.
\newblock \bibinfo{title}{{PROV-Overview: An Overview of the PROV Family of
  Documents}}.
\newblock
\newblock
\urldef\tempurl%
\url{https://www.w3.org/TR/prov-overview/}
\showURL{%
\tempurl}


\bibitem[\protect\citeauthoryear{Wang, Yu, Zheng, Gan, Gai, Ye, Li, Zhou,
  Huang, Ma, Huang, Guo, Zhang, Lin, Zhao, Li, Smola, and Zhang}{Wang
  et~al\mbox{.}}{2019b}]%
        {dgl}
\bibfield{author}{\bibinfo{person}{Minjie Wang}, \bibinfo{person}{Lingfan Yu},
  \bibinfo{person}{Da Zheng}, \bibinfo{person}{Quan Gan}, \bibinfo{person}{Yu
  Gai}, \bibinfo{person}{Zihao Ye}, \bibinfo{person}{Mufei Li},
  \bibinfo{person}{Jinjing Zhou}, \bibinfo{person}{Qi Huang},
  \bibinfo{person}{Chao Ma}, \bibinfo{person}{Ziyue Huang},
  \bibinfo{person}{Qipeng Guo}, \bibinfo{person}{Hao Zhang},
  \bibinfo{person}{Haibin Lin}, \bibinfo{person}{Junbo Zhao},
  \bibinfo{person}{Jinyang Li}, \bibinfo{person}{Alexander~J. Smola}, {and}
  \bibinfo{person}{Zheng Zhang}.} \bibinfo{year}{2019}\natexlab{b}.
\newblock \showarticletitle{Deep Graph Library: Towards Efficient and Scalable
  Deep Learning on Graphs}.
\newblock \bibinfo{journal}{\emph{CoRR}}  \bibinfo{volume}{abs/1909.01315}
  (\bibinfo{year}{2019}).
\newblock
\showeprint[arXiv]{1909.01315}
\urldef\tempurl%
\url{http://arxiv.org/abs/1909.01315}
\showURL{%
\tempurl}


\bibitem[\protect\citeauthoryear{Wang, Ji, Shi, Wang, Ye, Cui, and Yu}{Wang
  et~al\mbox{.}}{2019a}]%
        {han}
\bibfield{author}{\bibinfo{person}{Xiao Wang}, \bibinfo{person}{Houye Ji},
  \bibinfo{person}{Chuan Shi}, \bibinfo{person}{Bai Wang},
  \bibinfo{person}{Yanfang Ye}, \bibinfo{person}{Peng Cui}, {and}
  \bibinfo{person}{Philip~S Yu}.} \bibinfo{year}{2019}\natexlab{a}.
\newblock \showarticletitle{Heterogeneous graph attention network}. In
  \bibinfo{booktitle}{\emph{WWW}}. \bibinfo{pages}{2022--2032}.
\newblock


\bibitem[\protect\citeauthoryear{Xu, Hu, Leskovec, and Jegelka}{Xu
  et~al\mbox{.}}{2019}]%
        {gin}
\bibfield{author}{\bibinfo{person}{Keyulu Xu}, \bibinfo{person}{Weihua Hu},
  \bibinfo{person}{Jure Leskovec}, {and} \bibinfo{person}{Stefanie Jegelka}.}
  \bibinfo{year}{2019}\natexlab{}.
\newblock \showarticletitle{How Powerful are Graph Neural Networks?}. In
  \bibinfo{booktitle}{\emph{International Conference on Learning
  Representations}}.
\newblock


\end{thebibliography}

\blankpage
\blankpage
\appendix
\section{Appendix}
\subsection{Application Level Sample Provenance Graphs}

\begin{figure} [!htb]
\includegraphics[width=6cm, height=6cm, keepaspectratio]{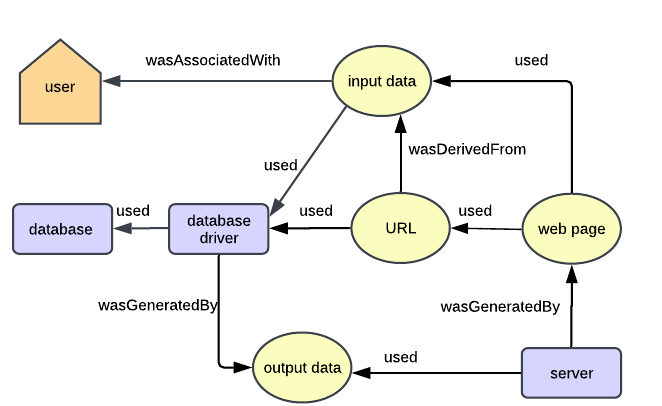}
\caption{Provenance graph for SQL injection attack.}
\label{fig:sqli}
\end{figure}

\begin{center}
\begin{figure} [!htb]
\includegraphics[width=6cm, height=6cm, keepaspectratio]{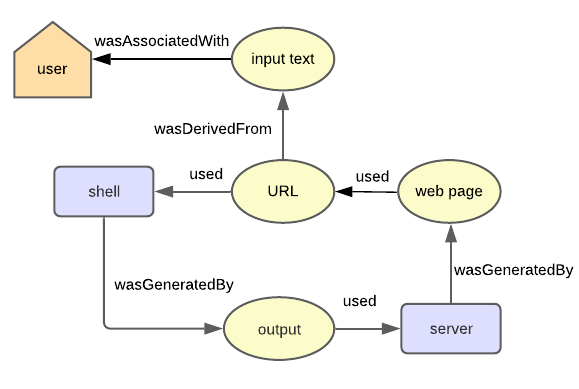}
\caption{Provenance graph for command line injection attack.}
\label{fig:cmdi}
\end{figure}

\begin{figure} [!htb]
\includegraphics[width=5cm, height=6cm, keepaspectratio]{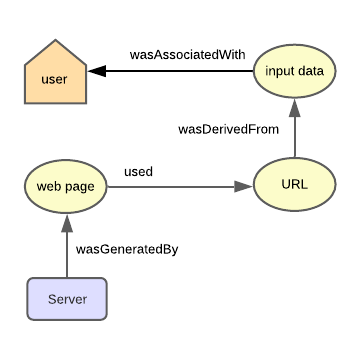}
\caption{Provenance graph for cross site scripting attack.}
\label{fig:xss}
\end{figure}
\end{center}

\subsection{Implementation Details}

The experiments were conducted on a 64\-core AMD EPYC 7742 CPU @ 2.25 GHz with an NVIDIA A100 GPU. Our models are implemented using PyTorch 1.10.2 and DGL 0.7.2 in Python 3.9. We use our Trainer and TrainingArguments classes to specify and define training and evaluation behavior, and we use DGL to implement our message passing graph neural network models. Evaluation metrics are computed using scikit\-learn \textit{f1\_score}, \textit{prescision\_score}, and \textit{recall\_score}. Experiments were conducted using five-fold cross validation with metrics reported as the mean and standard deviation averaged over the five folds.

\paragraph{Parameter Configuration.} We set the hidden and embedding dimensions for all models to 256 dimensions for all experiments, and the attention dimension for HAN and PROV-GEm are set to 64 dimensions. A dropout out rate of 0.5 and L2 regularization weight decay of 0.005 were used as regularization during model training. Table~\ref{tab:params} defines the individual training parameters used for each model for each experiment with the six attack-type datasets generated by \textsc{Flurry}.

\subsection{Learning Parameters}
\begin{table} [hbt!]
\begin{tabular}{lllllll|}
\hline
Attack Type   & Model    & Epochs & Aggregation & Learn Rate \\ \hline
              & HAN      & 20     & sum           & 0.001          \\
brute-force   & R-GCN    & 10     & sum           & 0.001         \\
              & PROV-GEm & 20     & sum           & 0.001         \\ \hline
              & HAN      & 7      & sum           & 0.0005               \\
cl-injection  & R-GCN    & 7      & sum           & 0.0005 \\
              & PROV-GEm & 7      & sum           & 0.0005 \\ \hline
              & HAN      & 5      & sum           & 0.0005  \\
sql-injection & R-GCN    & 20     & mean          & 0.0005  \\
              & PROV-GEm & 10     & sum           & 0.0005  \\ \hline
              & HAN      & 20     & sum           & 0.001 \\
xss-dom       & R-GCN    & 10     & sum           & 0.0005  \\
              & PROV-GEm & 20     & sum           & 0.001 \\ \hline
              & HAN      & 20     & sum           & 0.001  \\
xss-reflected & R-GCN    & 20     & mean          & 0.0005  \\
              & PROV-GEm & 20     & sum           & 0.001   \\ \hline
              & HAN      & 7      & sum           & 0.005  \\
xss-stored    & R-GCN    & 7      & sum           & 0.001  \\
              & PROV-GEm & 7      & sum           & 0.005   \\ \hline
\end{tabular}
\caption{Training Parameters for Graph Learning Models}
\label{tab:params}
\end{table}

\end{document}